\documentclass[reprint,
superscriptaddress,
showpacs,
preprintnumbers,
nofootinbib,
amsmath,
amssymb, 
aps,
prd,
floatfix
]{revtex4-1}

\usepackage[utf8]{inputenc}
\usepackage[normalem]{ulem}
\usepackage{graphicx}% Include figure files
\usepackage{dcolumn}% Align table columns on decimal point
\usepackage{bm}% bold math
\usepackage[colorlinks=true,allcolors=magenta]{hyperref}
\usepackage{url}

\usepackage{slashed,multirow,relsize,soul,tikz}
\usepackage{color}
\usepackage{mathrsfs} % pretty maths
\usepackage{amsmath} % pretty maths

\newcommand{\refeq}[1]{Eq.~(\ref{#1})}

\newcommand{\reffig}[1]{Fig.~\ref{#1}}

\newcommand{\reftab}[1]{Table~\ref{#1}}

\def\eg{\emph{e.g.}}

\newcommand{\mbeq}{\overset{!}{=}}

\newcounter{CommentCount}
\definecolor{PB}{rgb}{0.9,0,0}
\definecolor{MH}{rgb}{0.0,0.9,0}
\definecolor{SP}{rgb}{0.0,0.0,0.9}
\definecolor{palatinate}{rgb}{0.494, 0.192, 0.482}
\begin{document}

\preprint{\hfill IPPP/19/19/FTPI-MINN-20-17}

\title{Dark neutrinos and a three portal connection to the Standard Model}% 

\author{Peter Ballett}
\affiliation{Institute for Particle Physics Phenomenology, Department of
Physics, Durham University, South Road, Durham DH1 3LE, United Kingdom.}

\author{Matheus Hostert}
\email{mhostert@umn.edu}
\affiliation{School of Physics and Astronomy, University of Minnesota, Minneapolis, MN 55455, USA}
\affiliation{William I. Fine Theoretical Physics Institute, School of Physics and Astronomy, University of
Minnesota, Minneapolis, MN 55455, USA}
\affiliation{Perimeter Institute for Theoretical Physics, Waterloo, ON N2J 2W9, Canada}

\author{Silvia Pascoli}
\email{silvia.pascoli@durham.ac.uk}
\affiliation{Institute for Particle Physics Phenomenology, Department of
Physics, Durham University, South Road, Durham DH1 3LE, United Kingdom.}

\date{\today}

\begin{abstract}
We introduce a dark neutrino sector which respects a hidden $U(1)^\prime$ gauge symmetry, subsequently broken by the vacuum expectation value of a dark scalar. The model is a self-consistent realisation of an extended hidden sector that communicates with the SM only via the three renormalizable portals, namely neutrino, vector and scalar mixing. The interplay between portal couplings leads to several novel signatures in heavy neutrino, dark photon, and dark scalar searches, typically characterised by multi-leptons plus missing energy and displaced vertices. A striking signature arises in kaon factories such as NA62, where $K^+\to\ell_\alpha^+\nu \ell_\beta^+\ell_\beta^-$ decays could reveal a heavy neutrino and a light dark photon resonance above backgrounds. Given the open parameter space, we also comment on recent ideas to explain outstanding experimental anomalies, and how they would fit in our proposed model. A minimal extension of the model, possibly motivated by anomaly cancellation, can accommodate a dark matter candidate strongly connected to the neutrino sector.
\end{abstract}

\maketitle

%%%%%%%%%%%%%%%%%%%%%%%%%%%%%%%%%%%%%%%%%%%%%%%%%%%%%%%%%%%%%%%%%%%%%%%
\section{Introduction} 
% DIAGRAMS

The most important evidences that the Standard Model (SM) of particle physics is incomplete are neutrino masses and mixing, and the presence of dark matter (DM) in the Universe. Both call for extensions of the SM and the possible existence of dark sectors which do not partake in SM interactions, or do so with extremely weak couplings while displaying strong ``dark" interactions~\cite{Boehm:2003hm,Boehm:2003ha,Alexander:2016aln}.
Such sectors might exist at relatively light scales below the electroweak one, being within reach of present and future non-collider experiments. Generically, a neutral dark sector can communicate with the SM via three renormalizable portals. New neutral fermions mix with light neutrinos unless a symmetry differentiates the two, a possibility usually denoted as the neutrino portal. New vector particles can kinetically mix with the SM hypercharge, and new scalars mix with the Higgs boson through the so-called vector and scalar portals, respectively. The latter terms are generically allowed in the Lagrangian and an explanation of their smallness requires specific UV completions.  

In this article, we propose a new neutrino model with a hidden $U(1)^\prime$ gauge symmetry under which no SM fields are charged. We introduce new SM-neutral fermions, $\nu_D$ and an additional sterile neutrino $N$. The symmetry is subsequently broken by the vacuum expectation value (vev) of a complex dark scalar $\Phi$, which gives mass to the new gauge boson. For concreteness, we restrict the scale of the breaking to be below the electroweak one. 
\begin{figure}[b]
    \centering
\includegraphics[width=0.35\textwidth]{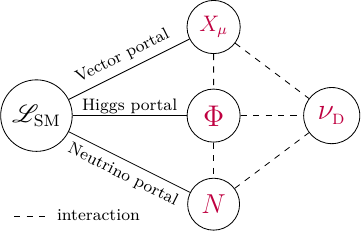}
\caption{Schematic representation of our three-portal model.\label{fig:model_diagram}}
\end{figure}

Models with heavy neutrinos which are not completely sterile and might participate in new gauge interactions have been studied in several contexts, including $B-L$, $L_\mu-L_\tau$, and left-right symmetric models~\cite{%
Buchmuller:1991ce,% original
Khalil:2006yi,% B-L at TeV scale
Perez:2009mu,% B-L extensive discussion + X model
Khalil:2010iu,% B-L at collider and ISS
Dib:2014fua,% B-L in linear seesaw
Baek:2015mna,% mu-tau in inverse seesaw
DeRomeri:2017oxa,% Elusive B-L
Nomura:2018mwr,% U(1)_R
Brdar:2018sbk% LR symmetry low scale
}, but here we focus on the possibility of a symmetry under which no SM fields are charged~\cite{%
Okada:2014nsa,% U(1)prime
Diaz:2017edh,% Hidden with loops
Nomura:2018ibs,% Hidden in linear seesaw
Hagedorn:2018spx,% Dark scotogenic
Shakya:2018qzg}. New heavy neutral fermions that feel such hidden forces, such as $\nu_D$, are referred to as \textit{dark neutrinos}, since they define a dark sector separate from the SM. Nevertheless, the dark interactions ``leak" into the SM sector via neutrino mixing, where they may dominate~\cite{Pospelov:2011ha,Batell:2016zod}. Models of this type have been invoked to generate large neutrino non-standard interactions~\cite{Farzan:2015doa,Farzan:2016wym}, generate new signals in DM experiments~\cite{Pospelov:2011ha,Pospelov:2012gm,Pospelov:2013rha,Harnik:2012ni,McKeen:2018pbb}, weaken cosmological and terrestrial bounds on eV scale sterile neutrinos~\cite{Hannestad:2013ana,Dasgupta:2013zpn,Mirizzi:2014ama,Chu:2015ipa,Cherry:2016jol,Chu:2018gxk,Denton:2018dqq,Esmaili:2018qzu}, and as a potential explanation of anomalous short-baseline results at the MiniBooNE~\cite{AguilarArevalo:2007it,Aguilar-Arevalo:2018gpe} and/or LSND~\cite{PhysRevLett.77.3082,Aguilar:2001ty} experiments with new degrees of freedom at the MeV/GeV scale~\cite{Gninenko:2009ks,Gninenko:2010pr,Masip:2012ke,Radionov:2013mca,Ballett:2018ynz,Bertuzzo:2018itn,Arguelles:2018mtc}.
% % %

Our model presents all the three renormalizable portals to the SM in a self-consistent way. The Yukawa interactions between the leptonic doublet and $N$, and between $N$ and $\nu_D$ induce neutrino mixing. 
The gauge symmetry allows a cross-coupling term in the potential between the Higgs and the real part of the scalar, inducing mixing between the two after symmetry breaking. The broken gauge symmetry implies the existence of a light hidden gauge boson $X_\mu$, which mediates the dark neutrino interactions and generically kinetically mixes with the SM hypercharge. This self-consistent setup combines the three portals into a unified picture that exhibits significantly different phenomenology with respect to each portal taken separately. Of particular interest is the fact that the different portal degrees of freedom display novel decay channels and scattering properties, and in many cases would have escaped experimental searches performed to date. We provide a selective list of the most affected bounds, and highlight the most interesting novel signatures that arise. In particular, we propose a new experimental search for the simultaneous presence of heavy neutrino and vector resonances in rare leptonic meson decays. In view of the relatively unexplored parameter space of our model, we comment on how these particles can explain long-standing experimental anomalies. We focus on a novel explanation to the MiniBooNE anomaly, based on the discussion of Ref.~\cite{Ballett:2018ynz} (see also Refs.~\cite{Bertuzzo:2018itn,Arguelles:2018mtc}), where new neutrino scattering signatures arise at neutrino experiments. We also reconsider the possibility to explain the discrepancy between the prediction~\cite{Davier:2010nc,Davier:2017zfy,Blum:2018mom,Keshavarzi:2018mgv,Davier:2019can} and measurement~\cite{Bennett:2006fi} of the anomalous magnetic moment of the muon ($\Delta a_\mu$) via kinetic mixing~\cite{Fayet:2007ua,Pospelov:2008zw}. Finally, we comment on how a scalar that couples strongly to dark neutrinos can help explain the anomalous $K_L\to\pi^0\nu\overline{\nu}$ events at KOTO~\cite{KOTO:2019}.

An interesting feature of the model is the generation of neutrino masses at loop-level. This requires only two key features of our setup, namely a light $Z^\prime$ and neutrino mixing, but not the vector and scalar portals. For this reason, we discuss it elsewhere~\cite{Ballett:2019cqp}.

In its minimal form, the model is not anomaly-free. We discuss how this can be cured and propose a minor extension that introduces additional dark sector neutral fermions charged under the new symmetry~\cite{Boehm:2003hm,Boehm:2003ha}. Neutrinos, we argue, may be a window into such dark sectors, bridging the puzzles of neutrino masses and DM~\cite{Ma:2006km,Farzan:2009ji,Farzan:2010mr,Arhrib:2015dez,Cherry:2014xra,Escudero:2016tzx,Escudero:2016ksa,Batell:2017cmf,Capozzi:2017auw,Campo:2017nwh,Blennow:2019fhy}. We briefly outline the key features of a DM extension and leave a more detailed analysis to future work.

%%%%%%%%%%%%%%%%%%%%%%%%%%%%%%%%%%%%%%%%%%%%%%%%%%%%%%%%%%%%%%%%%%%%%%%
\section{The Model} 
We extend the SM gauge group with a new abelian gauge symmetry $U(1)^\prime$ with associated mediator $X_\mu$ and introduce three new singlets of the SM gauge group: a complex scalar $\Phi$, and two left-handed fermions $\nu_{D,L} \equiv \nu_{D}$ and $N_L \equiv N$. 
%As shown in \reftab{tab:fields}, t
The scalar $\Phi$ and the fermion $\nu_{D}$ are equally charged under the new symmetry, and $N$ is neutral with respect to all gauge symmetries of the model. For simplicity, we restrict our discussion to a single generation of hidden fermions. The relevant terms in the gauge-invariant Lagrangian are 
\begin{align} \label{eq:lagrangian}
\mathscr{L}  \supset  &\left(D_\mu \Phi\right)^\dagger \left(D^\mu \Phi\right) -  V(\Phi,H) \,   \nonumber\\
&  - \frac{1}{4}X^{\mu \nu} X_{\mu \nu} + \overline{N}i\slashed{\partial}N + \overline{\nu_D}i\slashed{D}\nu_D 
\nonumber\\
&- \left[y^\alpha_\nu (\overline{L_\alpha} \cdot \widetilde{H})N^c + \frac{\mu^\prime}{2}\overline{N}N^c + y_N \overline{N}\nu_D^c\Phi + \text{h.c.}\right],
\end{align}
where $X^{\mu\nu}$ is the field strength tensor for $X_{\mu}$, $D_\mu \equiv \left(\partial_\mu-ig^\prime X_\mu\right)$ the covariant derivative, $L_\alpha \equiv (\nu_\alpha^T, \ell_\alpha^T)^T$ the SM leptonic doublet of flavour $\alpha = e, \mu, \tau$ and $\widetilde{H} \equiv i \sigma_2 H^*$ is the charge conjugate of the SM Higgs doublet. We write $y_\nu^\alpha$ for the $L_\alpha$--$N$ Yukawa coupling, $y_N$ for the $\nu_D$--$N$ one, and $\mu^\prime$ for the Majorana mass of $N$, which is allowed by the SM and the new gauge interaction, although it breaks lepton number by 2 units.

The minimisation of the scalar potential $V(\Phi,H)$ leads the neutral component of the fields $H$ and $\Phi$ to acquire vevs $v_H$ and $v_\varphi$, respectively. The latter also generates a mass for both the new gauge boson $X_\mu$ and the real component of the scalar field $\varphi$. Although $v_\varphi$ is arbitrary, we choose it to be below the electroweak scale, $v_\varphi < v_H$, as we are interested in building a model testable at low scales.

\paragraph{Neutrino portal}
In the neutral fermion sector and after symmetry breaking, two Dirac mass terms are induced with $m_D \equiv y_\nu^\alpha v_H/\sqrt{2}$ and $\Lambda \equiv y_N v_\varphi/\sqrt{2}$.
It is useful to consider the form of the neutrino mass matrix in the single generation case to clarify its main features. For one active neutrino $\nu_\alpha$ ($\alpha= e, \mu, \tau$), it reads
\begin{align} \label{eq:massmatrix}
\mathscr{L}_{\rm mass} \supset
\frac{1}{2}\left (\begin{matrix} \overline{\nu}_\alpha & \overline{N} &  \overline{\nu_D} \end{matrix} \right )
\left(\begin{matrix} 
     0   &  m_D        & 0 
\\ m_D &  \mu^\prime & \Lambda 
\\   0   &  \Lambda  & 0
\end{matrix}\right)
\left (\begin{matrix} \nu_\alpha^c \\ N^c \\ \nu_D^c \end{matrix} \right) + {\rm h.c.}
\end{align}  
The form of this matrix appears in Inverse Seesaw (ISS)~\cite{Mohapatra:1986bd,GonzalezGarcia:1988rw} and in Extended Seesaw (ESS)~\cite{Barry:2011wb,Zhang:2011vh} models. In fact, it is the same matrix discussed in the so-called Minimal ISS~\cite{Dev:2012sg}, with the difference that in our case its structure is a consequence of the hidden symmetry.
After diagonalisation of the mass matrix, the two heavy neutrinos, $\nu_h$ with $h=4,5$, acquire masses. Assuming that $m_D \ll \Lambda$, we focus on two interesting limiting cases. 

In the \textit{ISS-like} limit, where $\Lambda \gg \mu^\prime$ and the two heavy neutrinos are nearly degenerate, we have 
\begin{align}
 m_5 \simeq - m_4 \simeq \Lambda ~,  \,\,  m_5-|m_4| = \mu^\prime ~,& \nonumber\,\,
 U_{\alpha 5} \simeq U_{\alpha 4} \simeq  \frac{m_D}{\sqrt{2}\Lambda}~, \\   U_{D i} \simeq \frac{m_D}{\Lambda}, \,\, U_{D5} \simeq U_{D4} \simeq \frac{1}{\sqrt{2}} ~, \,\, & U_{N5} \simeq U_{N4} \simeq \frac{1}{\sqrt{2}} ~.\nonumber
\end{align}

In the \textit{ESS-like} case, $\Lambda \ll \mu^\prime$, one heavy neutrino remains very heavy and mainly in the completely neutral direction $N$, and the other acquires a small mass via the seesaw mechanism in the hidden sector. We find
\begin{align}
m_4 \simeq -\frac{ \Lambda^2}{\mu^\prime}~, & \,\,  m_5 \simeq \mu^\prime ~, \,\, U_{\alpha 4} \simeq U_{\alpha 5}\sqrt{\frac{m_5}{\left|m_4\right|}}  \simeq  \frac{m_D}{\Lambda}~,\nonumber \\ \,\, U_{D i} \simeq \frac{m_D}{\Lambda},  &\,\,  U_{N5} \simeq U_{D4} \simeq 1 ~, \,\,  U_{D5} \simeq U_{N4} \simeq \frac{\Lambda}{\mu^\prime} ~.\nonumber
\end{align}
Note that to lower the scale of the mediators while maintaining the heavy neutrino masses large, one must lower the gauge and scalar quartic coupling. Finally, we note that the mixing parameters of $\nu_4$ and $\nu_5$ are correlated in our model. At tree-level, the relation
\begin{equation}
    \frac{|U_{a 5}|^2}{|U_{a 4}|^2} = \frac{m_4}{m_5},
\end{equation}
holds exactly for $a=\alpha,N,D$. Loop-corrections are expected to lead to small deviations from this relation. In practice, this implies that $\nu_4$ has typically larger active-heavy mixing than $\nu_5$.

The Yukawa terms in \refeq{eq:lagrangian} induce {\em neutrino mixing} between the active (light) and heavy (sterile, dark) neutrinos. In this model, similarly to the ISS and the ESS cases, this mixing can be much larger than the typical values required in type-I seesaw extensions to explain neutrino masses, making its phenomenology more interesting. The determinant of the mass matrix in \refeq{eq:massmatrix} is zero, and so light neutrino masses vanish at tree-level and do not constrain the values of the active-heavy mixing angles. This, however, is no longer the case at one-loop level, as light neutrino masses emerge through radiative corrections from diagrams involving the $\varphi^\prime$ and $Z^\prime$ particles, as well as SM bosons~\cite{Ballett:2019cqp}.

\paragraph{Scalar portal}
In the scalar potential, the symmetries of the model allow us to write down the following term
\begin{equation}
    V(\Phi,H) \supset \lambda_{\Phi H} \, H^\dagger H \left| \Phi\right|^2,
\end{equation}
where we identify $\lambda_{\Phi H}$ as the scalar portal coupling~\cite{Barger:2008jx}, responsible for mixing in the neutral scalar sector. If such a term exists, the scalar mass eigenstates $(h^\prime, \varphi^\prime)$ mix with the gauge eigenstates $(h, \varphi)$ as $h^\prime = h \cos{\theta}  - \varphi \sin{\theta} $ and  $\varphi^\prime = h \sin{\theta}  + \varphi \cos{\theta}$,
with a mixing angle $\theta$ defined by 
\begin{equation}
\tan{(2\theta)} \equiv \frac{\lambda_{\Phi H} v_{H} v_\varphi}{\lambda_H v_{H}^2 - \lambda_\varphi v_\varphi^2},
\end{equation}
where $\lambda_H$ and $\lambda_\varphi$ are the quartic couplings of the Higgs and $\Phi$ scalars, respectively. 

\paragraph{Vector portal}  Similarly, mixing also arises in the neutral vector boson sector from the allowed kinetic mixing term~\cite{Holdom:1985ag}
\begin{equation}
 \mathscr{L} \supset - \frac{\sin{\chi}}{2} \, F^{\mu \nu} X_{\mu \nu},
\end{equation} 
where $F_{\mu\nu}$ is the SM hypercharge field strength. This term may be removed with a field redefinition, resulting in three mass eigenstates $\left( A,\, Z^0,\, Z^\prime\right)$, corresponding to the photon, $Z^0$-boson and the hypothetical $Z^\prime$-boson. For a light $Z^\prime$, the $Z^\prime$ coupling to SM fermions $f$ to first order in the small parameter $\chi$ is given by
\begin{equation}
\mathscr{L} \supset - (e\,q_f\,c_{W}) \chi \,\overline{f} \gamma^\mu f\,Z^\prime_\mu ~,
\end{equation}
with  $q_f$ the fermion electric charge.

The values of $\chi$ and $\lambda_{\Phi H}$ are arbitrary and could be expected to be rather large. As such, we treat them as free parameters within their allowed ranges. Here, we merely note that with our current minimal matter content, $\chi$ and $\lambda_{\Phi H}$ receive contributions at loop level from the $(\overline{L}_\alpha \cdot \widetilde{H})N^c$ and $\overline{N} \nu_D^c \Phi$ terms, which are necessarily suppressed by neutrino mixing ($\chi \propto g^\prime e |U_{\alpha h}|^2$ and $\lambda_{\Phi H} \propto |U_{\alpha h}|^2$). These values constitute a lower bound and larger values should be expected in a complete model.

%%%%%%%%%%%%%%%%%%%%%%%%%%%%%%%%%%%%%%%%%%%%%%%%%%%%%%%%%%%%%%%%%%%%%%%
%%%%%%%%%%%%%%%%%%%%%%%%%%%%%%%%%%%%%%%%%%%%%%%%%%%%%%%%%
\section{Decay Rates} \label{sec:decay_rates}

The phenomenology of the model depends critically on the ordering of the heavy neutrinos and the dark bosons, which controls the decay channels and lifetimes of these particles. In what follows, we list the most relevant decay rates in our model, denoting by $\nu$ the combination of all light mass eigenstates ($\nu_1, \nu_2, \nu_3$) that can appear in a given process. For clarity and simplicity, we separate the light dark bosons and heavy dark bosons cases, defined by the condition $m_{Z^\prime,\varphi^\prime}<m_4$ or $m_4<m_{Z^\prime,\varphi^\prime}$, respectively. We ignore SM contributions to the decay rates, as these are typically sub-leading for the cases of interest.

\subsubsection{Heavy bosons}
For $Z'$ heavier than the heavy neutrinos, $m_{\nu_h}<m_{Z'}$, with $h=4,5$, and with large kinetic mixing, the heavy neutrinos decay predominantly via three-body decays with an off-shell boson. Unless specified, we assume that the mass of the dark scalar is heavier and does not contribute to the decay rates in this section. The SM $Z$ and the $Z^\prime$ contributions can interfere, although we are mainly interested in the case where the latter dominates. The decays of most interest are $\nu_5 \rightarrow \nu_4 \ell^+ \ell^- $ and $\nu_4 \rightarrow \nu \ell^+ \ell^-$, with $\ell=e, \mu$, as far as these channels are kinematically accessible. The decay length of $\nu_5$ critically depends on the mass difference between $\nu_5$ and $\nu_4$. For concreteness, we focus on specific benchmark points (BP) that illustrate the key features.
In the ISS-like regime, we take $m_4/m_5=99\%$ and choose $m_4\simeq m_5 = 100$~MeV. If $\chi$ is negligible, we have that $\nu_h$ decays as in the minimal sterile neutrino model case via SM interactions. This is because the $\nu_5\rightarrow \nu_4 \nu \bar{\nu}$ decay is phase-space suppressed ($\Gamma_{\nu_5 \to \nu_4 \nu\nu} \propto \mu^{\prime\, 5}$), and because $Z^\prime$ mediated decays into three light neutrinos are negligible for small mixing, as $\Gamma_{\nu_h \to \nu\nu\nu} \propto |U_{\alpha h}|^6 m_{\nu_h}^5/m_{Z^\prime}^4$, where $|U_{\alpha h}|$ is a small mixing parameter between $e$, $\mu$, and $\tau$ flavours with the heavy neutrinos. If $\chi$ is sizeable, on the other hand, new visible decay channels dominate, specifically $\nu_5 \rightarrow \nu e^+ e^-$ and $\nu_4 \rightarrow \nu e^+ e^-$ for this BP. In an ESS-like regime instead, where $m_4 = m_5/10$ for instance, $\nu_5$ decays into 3 $\nu_4$ states very rapidly. The subsequent decays of $\nu_4$ would proceed via neutrino and kinetic mixing, but would be much slower than the $\nu_5$ one given the hierarchy of masses and the further suppression due to the portal couplings. 

The most relevant decay rates for our case studies is that of heavy neutrinos into dilepton plus missing energy. Neglecting the electron mass, the decay rate of $\nu_5$ is given by 
\begin{align}\label{eq:threebodydecays}
    \Gamma(\nu_5 \to \nu_4 e^+ e^-) & \simeq |U_{D5}|^2 |U_{D4}|^2 \,F(\sqrt{x_4} ) \\\nonumber&\qquad\qquad\quad\times\frac{(e \,c_W \chi \,g^\prime)^2}{384 \pi^3} \frac{m_5^5}{m_{Z^\prime}^4},
\end{align}
where $x_4=m_4^2/m_5^2$ and $F(x)=1+2x-8x^2+18x^3-18x^5+8x^6-2x^7-x^8 +24 x^3(1-x+x^2)\log{x}$. Note that $F(0)=1$, and $F(1)=0$, with the function obtaining a maximum value of $F(0.09)\simeq1.09$. For $\sqrt{x_4}=1-\epsilon$, one can show that $F(x_4)\to\epsilon^5$ when $\epsilon\ll1$, indicating a strong suppression of this decay in ISS-like scenarios, as stated before. Both $\nu_4$ and $\nu_5$ may decay via 
\begin{align}
    \Gamma(\nu_h \to \nu e^+ e^-) & \simeq |U_{Dh}|^2(1-|U_{D4}|^2-|U_{D5}|^2) \\\nonumber&\qquad\qquad\quad\times\frac{(e \,c_W \chi \,g^\prime)^2}{384 \pi^3} \frac{m_h^5}{m_{Z^\prime}^4}.
\end{align}
For heavier masses, additional decay channels, e.g. $\nu_4 \rightarrow \nu \mu^+ \mu^-$ and $\nu_4 \rightarrow \nu \pi^+\pi^-$, would open. A feature of the model is that $\nu_4 \rightarrow \nu \mu^+ \mu^-$ would have the same BR as the $e^+e^-$ one, albeit phase space suppressed. Two-body decays into neutral pseudoscalars via the new force are heavily suppressed due to the vector nature of the gauge coupling, unless mass mixing with the $Z$ is introduced (see~\cite{Ilten:2018crw} for a thorough discussion of the decay products of a dark photon). Decays into vector mesons are enhanced, where, for instance, we find
\begin{align}
    \Gamma(\nu_h\to\nu\rho^0)&=|U_{Dh}|^2(1-|U_{D4}|^2-|U_{D5}|^2)
    \\
    \nonumber &\quad \frac{(e c_W \chi g^{\prime})^2}{16\pi}\frac{m_h^3 f_\rho^2}{m_{Z^\prime}^4}(1-r_\rho)^2\left(\frac{1}{2} +r_\rho\right),
\end{align}
for sufficiently heavy dark photons, and where $r_\rho = m_\rho^2/m_h^2$ and $f_\rho \simeq 210$ MeV.

The dominant decays of the heavy dark photon and dark scalar are into heavy neutrinos, and the relevant rates can be obtained from
\begin{align}
\Gamma(Z^\prime\to \nu_4 \nu_5) &=\nonumber |U_{D5}|^2 |U_{D4}|^2 \,\frac{g^{\prime2} m_{Z^\prime}}{12\pi}
\\
&\quad\times\left(1+\frac{\Delta r}{2}\right) \left(1-R\right)^{3/2} \sqrt{1-\Delta r},
\end{align}
where $R=(m_4+m_5)^2/m_{Z^\prime}^2$ and $\Delta r = (m_5-m_4)^2/m_{Z^\prime}^2$, and
\begin{align}
    \Gamma (\varphi^\prime \to \nu_4 \nu_5) &= |U_{D 5} U_{N 4}+U_{D 4} U_{N 5}|^2 \frac{y_N^2 m_{\varphi^\prime}}{16 \pi} \\\nonumber &\qquad\quad \times\left(1-R^\prime\right)^{3/2}\sqrt{1-\Delta r^{\prime}},
\end{align}
where $R^\prime=(m_{4}+m_{5})^2/m_{\varphi^\prime}^2$ and $\Delta r^\prime =(m_{5}-m_{4})^2/m_{\varphi^\prime}^2$. Note that as a consequence of $U_{Ni}=0$, for $i=1,2,3$, the decay $\varphi^\prime \to \nu\overline{\nu}$ vanishes at tree level, and can be neglected here.

\subsubsection{Light bosons}
If $m_{Z'}< m_{\nu_h}$, the heavy neutrino will dominantly and immediately decay as $\nu_h \rightarrow \nu_\alpha Z'$, or even faster for $\nu_5 \rightarrow \nu_4 Z'$ if this is kinematically allowed. For instance, the former decay rate is given by
\begin{align}
    \Gamma(\nu_h\to \nu Z^\prime) &= |U_{D h}|^2(1-|U_{D4}|^2-|U_{D5}|^2) \\\nonumber&\qquad\times\frac{g^{\prime 2}}{8\pi} \frac{m_{\nu_h}^3}{m_{Z^\prime}^2}\left(1-r\right)^2\left(\frac{1}{2}+r\right)~,
    \end{align}
where $r=m_{Z^\prime}^2/m_{\nu_h}^2$, which agrees with the rate in Ref.~\cite{Atre:2009rg} under the substitutions $Z^\prime\to Z$ and $g^\prime\to g/2 c_W$, but differs from the rate in Ref~\cite{Bertuzzo:2018itn}. For (pseudo-)Dirac $\nu_h$, the rate is smaller by a factor of two. If the scalar is light, $m_{\varphi^\prime} < m_{\nu_h}$, decay into the scalar degree of freedom is also possible, with 
\begin{align}
    \Gamma(\nu_h\to \nu \varphi^\prime) &= |U_{N h}|^2(1-|U_{D4}|^2-|U_{D5}|^2) \\\nonumber&\qquad\times\frac{y_N^{2}}{16\pi} \frac{m_{\nu_h}^3}{m_{\phi^\prime}^2}\left(1-r\right)^2\left(\frac{1}{2}+r\right)~,
    \end{align}
where $r=m_{Z^\prime}^2/m_{\nu_h}^2$.
The $Z'$ subsequently decays into $e^+ e^-$ via kinetic mixing with a decay rate
\begin{align}    
    \Gamma(Z^\prime\to \ell^+\ell^-) &= \frac{(e c_W\chi)^2}{12 \pi}m_{Z^\prime}\left(1+2 r_\ell\right) \sqrt{1-4 r_\ell},
\end{align}
where $r_\ell = m_\ell^2/m_{Z^\prime}^2$. An analogous expression holds for $\varphi^\prime\to \ell^+\ell^-$ decays, which are long lived due to the $m_\ell^2/v_H^2$ suppression from the Higgs Yukawa coupling.  It is given by
\begin{align}
    \Gamma(\varphi^\prime\to \ell^+\ell^-) &= \frac{ \sin^2{\theta}\,\, G_F}{\sqrt{2}} \frac{m_{\varphi^\prime}m_\ell^2}{4\pi} \left(1-4 r_\ell^\prime \right)^{3/2},
\end{align}
where $r_\ell^\prime=m_\ell^2/m_{\phi^\prime}^2$.

%%%%%%%%%%%%%%%%%%%%%%%%%%%%%%%%%%%%
%%%%%%%%%%%%%%%%%%%%%%%%%%%%%%%%%%%
\section{Three portal phenomenology}
\label{sec:portal_pheno}

We now present the most relevant signatures arising from the interplay between portal couplings and the heavy neutrinos $\nu_h$ ($h=4,5$). We begin with a discussion of a light dark photon scenario, where we present a novel search for exotic meson decays that can be performed at NA62. Later, we comment on how the usual searches for each particle changes with respect to the minimal models, separating our discussion in heavy neutrino, dark photon, and dark scalar searches. For each particle, we then point out how their modified behaviour in a three-portal model would fit in recently-proposed explanations to outstanding experimental anomalies. 
\begin{figure}[t]
    \centering
    \includegraphics[width=0.30\textwidth]{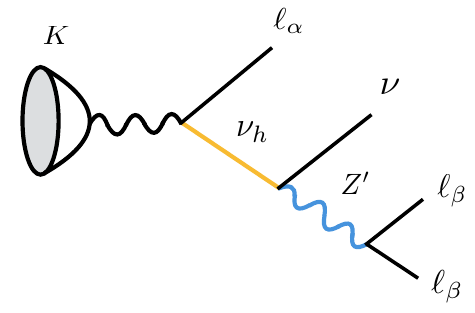}
    \caption{Kaon decay to a heavy neutrino that decays visibly either through a sequence of two-body decays or via a three-body decay.\label{fig:k_decay}}
\end{figure}

\begin{figure*}[t]
    \centering
  \includegraphics[width=0.49\textwidth]{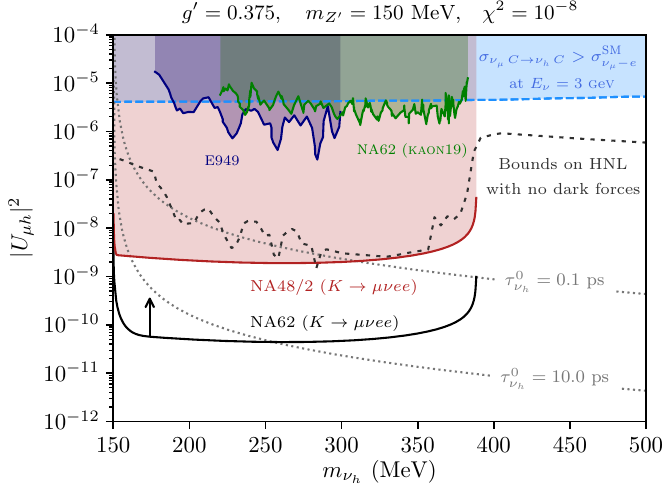}
  \includegraphics[width=0.49\textwidth]{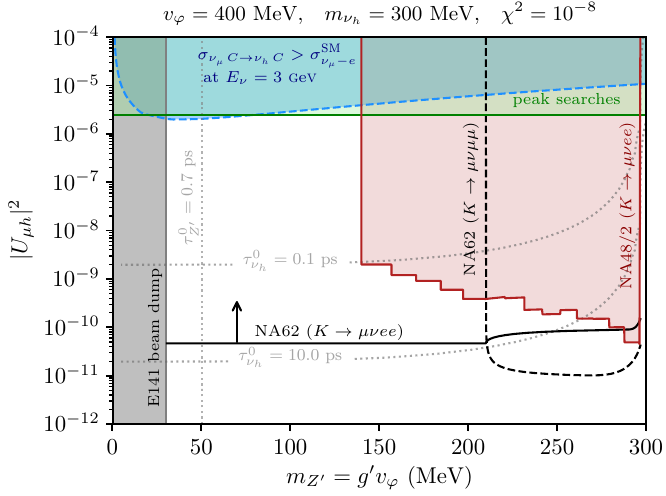}
    \caption{Parameter space in a phenomenological ISS scenario ($\mu^\prime\to0$) where a pseudo-Dirac pair decays visibly via $\nu_h\to\nu Z^\prime\to\nu e^+e^-$. On the left, we fix $m_{Z^\prime}$ and vary $m_{\nu_h}$, while on the left, we fix $m_{\nu_h}$ and vary $m_{Z^\prime}$. The NA62 single event sensitivity (SES) to $K^+\to\mu^+e^+e^-\nu$ and $K^+\to\mu^+\mu^+\mu^-\nu$ is shown as a solid and dashed black line, respectively. The resonances in $m_{\ell \ell}=m_{Z^\prime}$ and $m_{p_K-p_\mu}=m_{\nu_h}$ are expected to greatly reduce backgrounds, although the region $m_{ee} < 140$ MeV can still be challenging due to the large number of $\pi^0$ Dalitz decays. In red we show bounds obtained from the NA48/2 measurement of the rare leptonic kaon decay $K^+\to\mu\nu e^+e^-$~\cite{Peruzzo:2017qis}. Peak-search constraints on $\nu_h$ become less effective due to the fast visible decays, and we re-scale these assuming a conservative $0.5\%$ $e^+e^-$ detection inefficiency.
    In light blue we show the region where the upscattering cross section is larger than that of neutrino-electron scattering, implying that $e^+e^-$ pairs could be searched for in accelerator neutrino scattering experiments, such as MINER$\nu$A~\cite{Arguelles:2018mtc}. Beam dump constraints on $\nu_h$ disappear, but those on a light $Z^\prime$ remain (for $\chi^2=10^{-8}$, E141~\cite{Riordan:1987aw} provides the strongest bound). The dotted grey contours show a fixed total lifetime for $\nu_h$ and $Z^\prime$. For $|U_{\mu h}|>10^{-4}$, invisible $Z^\prime\to \nu \overline{\nu}$ decays start to dominate. }
    \label{fig:peak_searches}
\end{figure*}
\subsection{Exotic meson decays} From the discussion above, it is clear that if the dark bosons are lighter than the heavy neutrinos, then $\nu_h$ decays are fast, and become visible whenever the decays of the light bosons are fast and visible. Because the light scalars typically decay with much longer lifetimes than the dark photons, we will always assume that it lies at much larger masses and will not play a role in our study, unless explicitly stated otherwise. As a proof of principle, we focus on a scenario with a single pseudo-Dirac heavy neutrino $\nu_h$ in an ISS-like regime ($\mu^\prime\to0$), although it should be noted that in this case neutrino masses vanish. The case of Majorana neutrinos that contribute to light neutrino masses is completely analogous, noting that the mass mechanism is mostly insensitive to small kinetic and scalar mixing parameters. For a comparison with the relevant region for neutrino masses, see Ref~\cite{Ballett:2019cqp}. 

The most striking signatures associated with the light dark photon scenario arise in kaon factories, where meson decays to heavy neutrinos lead to visible signatures inside the detector. In this context, we propose a dedicated search for the following process
\begin{align}
    M^+ \,\to\, \ell_\alpha^+ \,\, \nu_h \to \ell_\alpha^+ \,\,\nu \,\, Z^\prime\,\to\, \ell_\alpha^+ \,\, \nu\,\,\ell^+_\beta\,\,\ell^-_\beta ,
\end{align}%
with $\alpha,\beta \in \{e, \mu\}$ (see \reffig{fig:k_decay}). For prompt dark cascades, the BR for such decays is simply BR$(M^+\to \ell_\alpha^+ \nu_h)$ = $|U_{\alpha h}|^2 \rho_\alpha(m_{\nu_h})$ BR$(M^+\to \ell_\alpha^+ \nu_\alpha)$, where $\rho_\alpha(m_{\nu_h})$ is the Shrock function~\cite{Shrock:1980ct}, which accounts for the heavy neutrino mass in such meson decays. This signature can be searched for by applying a simultaneous requirement of
\begin{align}
    m_{\beta \beta}^2 &\equiv (p_{\beta^+} + p_{\beta^-})^2 \mbeq m_{Z^\prime}^2,\\
    m_{p_K - p_\alpha}^2 &\equiv(p_K - p_{\alpha})^2 \mbeq m_{\nu_h}^2,\\
    m_{\rm miss}^2 &\equiv(p_K-p_{\alpha} -p_{\beta^+}-p_{\beta^-})^2 \mbeq 0,
\end{align}
within detector resolution. In addition, the strong correlation and smooth distribution over the invariant masses $m_{\alpha\nu}^2\equiv(p_K-p_{\beta^+}-p_{\beta^-})^2$ and $m_{\alpha\beta\beta}^2\equiv(p_{\alpha}+p_{\beta^+}+p_{\beta^-})^2$ can be used to further reduce backgrounds. From kinematics alone, it is possible to show that
\begin{equation}\label{eq:mass_relation}
    m_{\alpha \beta \beta }^2 + m^2_{\alpha \nu} + m_{p_K-p_\alpha}^2= m_K^2 +m_{\alpha}^2 + m_{\rm miss}^2 + m_{\beta\beta}^2,  \
\end{equation}
so that a measurement of one can be tested against the other for each point in parameter space.

\renewcommand{\arraystretch}{1.2}
\begin{table}[t]
    \centering
    \begin{tabular}{|c|c|c|}
        \hline &  $K^+\to \mu^+\nu e^+ e^-$ & $K^+\to \mu^+\nu \mu^+\mu^-$\\\hline\hline
        $N_K^{\rm Fid}$ & $2.14\times10^{11}$ & $7.94\times 10^{11}$\\
        ${\rm A}_\beta$ & $4\%$ & $10\%$ \\\hline
    \end{tabular}
    \caption{Assumptions used for the computation of NA62 single event sensitivity for heavy neutrinos decaying via two-body cascades into two charged leptons plus missing energy.}
    \label{tab:NA62values}
\end{table}
Ultimately, the degree of background reduction is highly detector-dependent as it arises mainly from resolution and particle identification effects. We return to this issue below, but a detailed analysis is left to the sophisticated detector simulations of the experimental collaborations. We do remark, however, that if the HNLs and the dark photon have lifetimes above $10$~ps in the rest frame, then they may lead to displaced vertices at NA62, as the boost factors are $\mathcal{O}(100)$ and the resolution for displaced vertices is of $\mathcal{O}(10)$ cm~\cite{EVGUENI}. This provides yet another tool to reduce backgrounds and is particularly useful in the small mixing region where the new physics events are expected to be small. Under the assumption of a light dark photon, a pseudo-Dirac pair that mixes only with the muon flavour, and neglecting final state masses, the typical lifetimes are
\begin{align}
    \tau^0_{\nu_h} &\simeq 1.7 \text{ ps } \times \left( \frac{10^{-10}}{|U_{D h}|^2|U_{\mu h}|^2}\right)\left( \frac{300 \text{ MeV}}{m_{\nu_h}}\right)^3\\&\qquad\qquad\qquad\times\left(\frac{0.375}{g^\prime}\right)^2 \left(\frac{m_{Z^\prime}}{100\text{ MeV}} \right)^2,
    \\
    \tau^0_{Z^\prime} &\simeq 0.70 \text{ ps }\times \left( \frac{10^{-4}}{\chi}\right)^2\left( \frac{50 \text{ MeV}}{m_{Z^\prime}}\right),
\end{align}
where we assumed BR$(\nu_h \to \nu Z^\prime)\simeq1$ and BR$(Z^\prime\to e^+e^-)\simeq 1$, with the understanding that $(|U_{D4}|^2+|U_{D5}|^2)(1-|U_{D4}|^2-|U_{D5}|^2) = |U_{D h}|^2|U_{\mu h}|^2$ as $\mu^\prime \to 0$. 

In what follows, we obtain an estimate of the single event sensitivity of NA62, assuming a zero-background search. We do this exercise for only $30\%$ of the collected dataset corresponding to the period between 2016 and 2018~\cite{CortinaGil:2019dnd}. Using $N_{K}$ useful kaon decays, down-scaled by a trigger factor to yield a $N_K^{\rm Fid}$ fiducial kaon decays, and a detector acceptance ${\rm A_\beta}$, we can find the experiment single event sensitivity to muon-heavy neutrino mixing in $K^+\to\mu^+\nu \ell_\beta \ell_\beta$ decays as
\begin{equation}
    |U_{\mu h}|^2 = \frac{1}{N_K^{\rm Fid} \,{\rm A_\beta}\,\, {\rm BR}(K\to\mu\nu_\mu)} \frac{1}{P_{\rm dec}},
\end{equation}
where
\begin{align}
P_{\rm dec} &\simeq \left(1-e^{\frac{-\langle L \rangle}{L_{\nu_h}}}\right)\,{\rm BR(\nu_h\to\nu Z^\prime)} \\\nonumber
&\qquad\qquad \times\left(1-e^{\frac{-\langle L \rangle}{L_{Z^\prime}}}\right)\,{\rm BR(Z^\prime \to \ell_\beta\ell_\beta)}    
\end{align}
is a crude approximation for the probability for $\nu_h$ and $Z^\prime$ to decay within the average distance between its production and the end of the detector, $\langle L \rangle$, when it has a decay length $L_P$ in the laboratory frame. For concreteness, we take $\langle L \rangle= 37.5$ m. At NA62, $\beta=e$ events ought to be registered by the ``di-electron'' and $\beta=\mu$ by the ``di-muon" trigger~\cite{CortinaGil:2019dnd}, each collecting only a fraction of the total useful kaon decays, and corresponding to a given $N_K^{\rm Fid}$ kaon decays in the fiducial volume as shown in \reftab{tab:NA62values}. We assume a constant acceptance ${\rm A}_{\beta}$ for each final state that is approximately the one obtained in searches for $K^+\to\pi^-\ell_\beta^+\ell_\beta^+$~\cite{CortinaGil:2019dnd} and $K^+\to\pi^+ (\pi^0\to\gamma(Z^\prime\to e^+e^-))$~\cite{CortinaGil:2019nuo} at NA62, also shown in \reftab{tab:NA62values}.

Our results are shown in \reffig{fig:peak_searches}. It is clear that within our assumptions of no backgrounds, NA62 can probe a large region of unexplored parameter space. Of most relevance is the region at low $m_{Z^\prime}$, where other bounds are rather weak. Incidentally, the di-electron search would directly test the model proposed in Ref.~\cite{Bertuzzo:2018itn} (see MiniBooNE discussion below), where a more phenomenological model with a single heavy neutrino is used. There, all dark couplings are fixed to be large, with $|U_{\mu 4}|^2 \gtrsim 10^{-9}$ and $m_4 \gtrsim 30$ MeV. Fixing $g^\prime$ to be $\mathcal{O}(1)$ implies that heavy neutrinos decay even faster, and the sensitivity curve in \reffig{fig:peak_searches} would apply as shown. In our figure, however, we choose to fix $v_\varphi$ so that $g^\prime$ must vary if $m_{Z^\prime}$ varies. If one were to fix $g^\prime$, only the upscattering cross section region in the plot would be affected, as the cross sections are proportional to $(g^\prime e \chi)^2|U_{\mu h}^*U_{D h}|^2$.

We do not study this case here, but $K^+\to e^+\nu\ell_\beta^+\ell_\alpha^-$ can also be searched for, and would lead to bounds on $|U_{e h}|^2$ mixing angle, which for promptly decaying Dirac neutrinos is also weakly constrained. Much heavier $Z^\prime$ particles imply heavy neutrino decay is a three-body process, and may not be so prompt. In that case, the resonance in $m_{\beta\beta}$ is no longer present, making background reduction more challenging. A peak search in $m_{p_K - p_\alpha}^2$ can still be performed, together with the requirement that the invariant masses measured obey \refeq{eq:mass_relation}, which is still valid.
The lifetimes of $\nu_h$ in the ISS-like regime of our model are too long to realise this signature, but in an ESS-like regime, $\nu_5 \to \nu_4 \ell_\beta^+\ell_\beta^-$ decays are sufficiently fast.

\paragraph{Backgrounds} We now discuss backgrounds to our proposed search. An irreducible but smooth background from SM radiative decays $M^+\to\ell_\alpha^+ \nu \ell_\beta^+ \ell_\beta^-$ exists at a BR of $\mathcal{O}(10^{-8})$, also displaying $m_{\rm miss}^2 =0$. Our signal, however, constitutes a single peak in the $m_{\beta\beta}$ vs $m_{p_{K}-p_{\alpha}}$ plane. Given that NA62 has already achieved an invariant mass resolution of $\delta m_{p_{K}-p_{\alpha}} = 1-20$ MeV~\cite{CortinaGil:2017mqf}, depending on $m_{p_{K}-p_{\alpha}}$, and that NA48/2 has achieved $\delta m_{ee}/m_{ee} \simeq 1\%$~\cite{Batley:2015lha}, we expect radiative leptonic kaon decays to not limit the sensitivity once \refeq{eq:mass_relation} has been taken into account. For $K^+\to\mu^+\nu e^+e^-$, the most challenging background appears at $m_{ee}<m_{\pi^0}$, where the large number of $K^+\to\mu^+\nu_\mu(\pi^0\to\gamma e^+e^-)$ decays can spoof our signature due to the soft nature of the photon in the pion Dalitz decay. The BR of such decays to fake our final states can be naively estimated to be around $10^{-8}$ for the excellent photon rejection at NA62 of $\simeq10^{-4}$. This, again, can be further reduced by enforcing invariant mass peaks and correlations. Similar considerations apply for any channel with Dalitz decays. Other backgrounds from $K^+\to (\pi^+\to\mu^+\nu) e^+e^-$, for instance, may be reduced with a cut on $m_{\mu\nu}>m_{\pi^+}$, at the cost of reducing signal acceptance. While our assumption of no backgrounds is optimistic at low $m_{Z^\prime}$ values, the region with $m_{Z^\prime}>140$ MeV has been studied before in the context of rare leptonic kaon decays, and presents more manageable backgrounds. For instance, the NA48/2~\cite{Peruzzo:2017qis} collaboration has performed a measurement of 
\begin{equation}
    {\rm BR} (K^+\to\mu^+\nu e^+e^-)  = (7.81 \,\pm\, 0.21\, {\rm stat.})\times 10^{-8}
\end{equation}
for $m_{ee}>140$ MeV, where we omitted the smaller systematic uncertainties. This measurement achieved an overall signal-to-background ratio of $\simeq 30$, with an estimate of 54 backgrounds events. 

For the dimuon channel, $K^+\to \pi^+\pi^+\pi^-$ presents the largest background rate, followed by $K^+\to \mu^+\nu\pi^+\pi^-$. These channels are challenging due to the subsequent decays of the $\pi^\pm\to\mu^\pm\nu$ as well as the mis-identification rate of $\pi\rightleftharpoons\mu$ of $0.4-0.9\%$ at NA62. The radiative leptonic decay $K^+\to \mu^+ \nu \mu^+ \mu^-$ would also present a background. No measurement of such SM decays exists but the stringiest limit comes from E787, where it was found that 
\begin{equation}
    {\rm BR}(K^+\to \mu^+ \nu \mu^+ \mu^-) < 4.7 \times 10 ^{-7}
\end{equation} 
at 90\% C.L. for $m_{\mu\mu} \in [220,320]$ MeV~\cite{Atiya:1989tq}. NA62 will measure such SM decays, as Ref.~\cite{Krnjaic:2019rsv} explores to set constraints on new light $L_\mu-L_\tau$ mediators. In comparison with the latter study, the multiple invariant mass resonances in our model would further reduce backgrounds and offer valuable insight if the light dark photon can decay to muons. A simultaneous detection in the electron and muon channels in accordance with the dark photon BRs would offer strong evidence for our kind of dark sector.

\paragraph{Similar measurements} A bound can already be derived from existing measurements compatible with the SM prediction in kaon radiative leptonic decays~\cite{Poblaguev:2002ug,Ma:2005iv,Peruzzo:2017qis}. The NA48/2 collaboration has measured the BR of $K^+\to\mu^+\nu e^+e^-$ as a function of $m_{ee}$, starting from $m_{ee}>140$ MeV, and great agreement between MC and data is observed~\cite{Peruzzo:2017qis}. This allows us to set constraints on $m_{Z^\prime}>140$ MeV, for various values of $m_{\nu_h}$. We show the region where the new physics events are larger than $20\%$ of the measured value for $m_{Z^\prime}=150$ MeV in the left panel of \reffig{fig:peak_searches}. The same is done for $m_{\nu_h}=300$ MeV and varying $m_{Z^\prime}$ in the left panel. Note that for $2 m_\mu<m_{Z^\prime}$, one would predict $K^+\to \mu^+ \nu \mu^+ \mu^-$ decay, which results in rather weak limit if compared with the $m_{ee}$-binned $K^+\to \mu^+ \nu e^+ e^-$ measurement at NA48/2.

Finally, similar signatures can arise in pion decays $\pi^+\to\ell^+_\alpha\nu e^+e^-$. For $\alpha=e$, a measurement compatible with the SM prediction was performed at the SINDRUM magnetic spectrometer~\cite{Egli:1986nk,Grab:1986si} where
\begin{equation}
    {\rm BR} \left(\pi^+ \to e^+ \nu e^+ e^-\right) = (3.4 \pm 0.5) \times 10^{-9}.
\end{equation}
No measurement or limit exists for $\pi^+\to\mu^+\nu e^+e^-$, where only $m_{\nu_h}\lesssim 34$ MeV heavy neutrinos can be tested. Production in muon~\cite{Bertl:1985mw} and tau~\cite{Alam:1995mt} leptonic decays of the type $\ell_\alpha\to \ell_\beta\ell_\gamma\ell_\gamma \nu \overline{\nu}$ with BR of $\mathcal{O}(10^{-5})$ are much less constraining, although tau decays would offer an unique process to test the less constrained $|U_{\tau h}|$ parameter.

\subsection{Impact on heavy neutrino searches} In this subsection and the ones that follow it, we will discuss a collection of interesting signatures and the most relevant changes to current experimental landscape in a model such ours, where at least two portal couplings may be at play. We do not make the assumption of a pseudo-Dirac HNL anymore, and use $\nu_h$ to denote both $\nu_4$ and $\nu_5$, whenever the distinction is not important. In presence of only neutrino mixing, the strongest bounds on heavy neutrinos in the MeV--GeV mass range come from peak searches in meson decays~\cite{
%KEK
Yamazaki:1984sj,
%NA48/2
Artamonov:2014urb,
%NA62
CERNNA48/2:2016tdo}
and beam dump experiments~\cite{
%PS-191
Bernardi:1985ny,
%CHARM
Bergsma:1983rt,
%NA3
Badier:1986xz,
%NuTeV
Vaitaitis:1999wq, 
%BEBC
CooperSarkar:1985nh, 
%NOMAD
Astier:2001ck} looking for visible $\nu_h$ decays. 
As we will see, both searches can be weakened if the $\nu_h$ decays are sufficiently different from the case of ``standard" sterile neutrinos with SM interactions suppressed by neutrino mixing.
 
\subsubsection{Peak searches and LNV} Peak searches in meson decays of the type $M\to\ell\nu_{h}$ have been long regarded as model-independent bounds on heavy neutrinos. This is due to the fact that only the parent meson and daughter charged lepton kinematics need to be measured in order to search for a peak in $(p_M - p_\ell)^2 = m_{\nu_h}^2$. We argue that the strict requirement of observing a single charged track in the detector~\cite{Artamonov:2014urb} would, however, veto a large fraction of new physics events if $\nu_h$ decays promptly into $\nu_\alpha \ell^+\ell^-$, for instance. This is not a concern in minimal sterile neutrino extensions of the SM due to the large lifetime of $\nu_h$, which is greatly reduced for large values of $\chi$ in our model. We illustrate this effect in the left panel of \reffig{fig:peak_searches}, where peak search bounds on $|U_{\mu h}|^2$ derived from $K^+\to\mu^+\nu_h$ decays at KEK~\cite{Yamazaki:1984sj}, E949~\cite{Artamonov:2014urb}, and NA62~\cite{CortinaGil:2017mqf,NA62HNL:2019} are reinterpreted in an ISS-like BP where a single heavy neutrino of Dirac nature undergoes 2-body decays into $\nu e^+e^-$ final states. With a probability to miss an $e^+e^-$ pair in the preceding detectors of $0.5\%$, this represents a 200 times weaker bound on the mixing angle. This inefficiency is clearly a conservative assumption based on the \emph{photon} detection inefficiency (typically larger than that of an $e^+e^-$ pair of the same energy) reported by the E949~\cite{Artamonov:2014urb} and NA62~\cite{NA62:2017rwk} collaborations. In addition, bounds on lepton number violation (LNV) from meson and tau decays are affected~\cite{Atre:2009rg,Abada:2017jjx}. These bounds are obtained by searching for $M^+\to \ell^+ \nu_h \to \ell^+ \ell^+ \pi^-$ decays, with the same-sign di-leptons being a smoking gun for LNV by two units. In our current model, the intermediate on-shell $\nu_h$ produced in said decays has very suppressed charged current branching ratios into $\ell^\pm \pi^\mp$ and $\ell^\pm K^\mp$ final states, and so such bounds are much weaker.

\subsubsection{Beam dump searches}

Beam dump and fixed target experiments are an ideal place to search for heavy neutrinos due to the large number of meson decays involved. The search strategy is based on producing such particles in decays at the target, and searching for their visible decay products in a detector located at a fixed distance from the target. If heavy neutrinos decay into visible particles faster than in the minimal sterile neutrino model, then such constraints on the mixing angle of heavy neutrinos are affected. As shown in Ref.~\cite{Ballett:2016opr}, if one only enhances the decay $\Gamma(\nu_h\to\nu e e)$ by a factor $\alpha$ with respect to its value in the minimal sterile model, then the upper bounds becomes stronger by a factor $\simeq \sqrt{\alpha}$, while the lower bounds becomes weaker by a factor of $\simeq \alpha$. This can be understood by noting that in the vicinity of the upper bound, the number of events is proportional to $|U_{\alpha h}|^4$, while in the lower bound it is proportional to $|U_{\alpha h}|^2$, as most particles decay inside the detector. For very short decay lengths, no bound can be placed, as all particles decay before reaching the detector. For this reason, beam dump bounds on $\nu_5$ typically vanish, while those on $\nu_4$ become stronger in an ESS-like regime. This typically introduces problems with searches at PS-191~\cite{Bernardi:1985ny}, but can be evaded in non-minimal realisations of our model where the correlation between the mixing parameters of $\nu_4$ and $\nu_5$ is broken. We come back to this issue when discussing MiniBooNE.

\subsection{Exotic neutrino scattering signatures} The presence of a light vector mediator and kinetic mixing can also induce new neutrino scattering signatures. For a hadronic target $\mathcal{H}$, light neutrinos may upscatter electromagnetically via the $Z^\prime$ into $\nu_h$, which subsequently decays into observable particles, \eg, $\nu_\alpha \, \mathcal{H} \to (\nu_h \to \nu\, \ell_\beta^+\,\ell_\beta^-)\, \mathcal{H}$. Beyond offering a novel scenario to explain the MiniBooNE low energy excess (see below), such upscattering can also produce exotic final states in neutrino detectors such as $e^+e^-$, $\mu^+\mu^-$, multi-meson, and $\tau^+\tau^-$ final states. If $m_{Z^\prime}<m_{\nu_h}$, the on-shell $Z^\prime$ decays can be searched for by looking for neutrino-produced di-leptons with $m_{\ell \ell} = m_{Z^\prime}$. Di-muons production, however, is tightly constrained due to kaon decay bounds (see above), unless $m_{\nu_h} \gtrsim 400$ MeV. In that case, signatures from $Z^\prime\to \pi^+\pi^-$ decays would appear for $2 m_{\pi^+}<m_{Z^\prime}$ MeV. If $m_{Z^\prime} \lesssim  140$ MeV, neutrino upscattering on nuclei is predominantly coherent, and leads exclusively to $e^+e^-$ final states. These can mimic single-photon showers in neutrino detectors, and could affect sideband measurements in neutrino-electron scattering data at (multi-)GeV energies (see Ref.~\cite{Arguelles:2018mtc}). Assuming an ISS-like model and a light dark photon, we show the region where the coherent upscattering cross section on Carbon ($\nu_\mu \,C\to \nu_h C$) is larger than that of neutrino-electron scattering in the SM in \reffig{fig:peak_searches}. When $m_{Z^\prime} \gtrsim 800$ MeV, upscattering happens predominantly on protons. In that case, more hadronic activity in the detector is expected, and neutrino-electron scattering measurements are no longer as effective. If $2m_\mu<m_{Z^\prime}<m_{\nu_h}$ or $2m_\mu<m_{\nu_h}<m_{Z^\prime}$, then such events contribute to neutrino-trident production ($\nu_\mu \, \mathcal{H} \to \nu_\mu \, \mu^+ \,\mu^- \, \mathcal{H}$) at neutrino scattering experiments if $\nu_h$ decays fast enough. Previous measurement at CCFR~\cite{Mishra:1991bv} and CHARM-II~\cite{Geiregat:1990gz} lead to weak bounds in the parameter space shown in \reffig{fig:peak_searches}, so we do not show them here. They become relevant when $m_{Z^\prime} \gtrsim 800$ MeV as neutrino-electron scattering becomes ineffective.

\subsubsection{MiniBooNE low energy excess} The upscattering signatures discussed above with $\ell^\pm = e^\pm$ have been invoked as an explanation of the excess of low energy electron-like events at MiniBooNE~\cite{Aguilar-Arevalo:2018gpe} in Refs.~\cite{Bertuzzo:2018itn} and ~\cite{Ballett:2018ynz}, where a simplified model containing a single heavy neutrino was used. This explanation relies on the fact that collimated or highly energy-asymmetric $e^+e^-$ pairs can mimic an electron-like signature in Cherenkov detectors, such as MiniBooNE. While the light dark photon case ($m_{Z^\prime}<m_{\nu_4}$)~\cite{Bertuzzo:2018itn} leads to tension with either the angular distribution at MiniBooNE or $\nu-e$ scattering data~\cite{Arguelles:2018mtc}, the heavy dark photon study ($m_{\nu_4}<m_{Z^\prime}$)~\cite{Ballett:2018ynz} finds a good fit to both energy and angular distributions at MiniBooNE, while evading constraints from $\nu-e$ scattering data. The preferred parameters were $m_4 = 140$ MeV, $m_{Z^\prime} = 1.25$ GeV, $g^\prime=1$, $|U_{\mu 4}|^2 = 1.5 \times 10^{-6}$, and $\chi^2 = 5\times10^{-6}$. There, the prompt decays of $\nu_4\to \nu e^+e^-$ were achieved by requiring a large mixing with the tau flavour, namely $|U_{\tau 4}|^2=7.8\times10^{-4}$. We note, however, that the decay lengths achieved by Ref.~\cite{Ballett:2018ynz} are too long when computed using our expressions, and may indicate that much larger values of dark couplings were used than the ones quoted. In addition, choosing values of $|U_{\tau 4}|^2 = 520 \times |U_{\mu 4}|^2$ or larger would imply that experiments with a large component of $\nu_\tau$ neutrinos in their beam would see $\nu_4$ production at a comparable rate to their total neutral-current elastic scattering rate. With a similar $e^+ e^-$-electron mis-identification invoked to explain the MiniBooNE excess, such events would lead to large numbers of $\nu_e$-like events at the far detector of T2K, in contradiction with $\nu_e$ appearance measurements~\footnote{We thank Pedro Machado for bringing this argument to our attention.}~\cite{Abe:2018wpn}. 

In an ESS-like limit of our current model, we predict that both $\nu_4$ and $\nu_5$ can be produced in upscattering. Due to the larger $\nu_5-\nu_4-Z^\prime$ coupling compared to that of $\nu_4-\nu_i-Z^\prime$, we can do away with the need of a large $|U_{\tau h}|^2$ mixing by producing sufficient numbers of $\nu_5$ states in the MiniBooNE detector. Take an analogous benchmark point to that of Ref.~\cite{Ballett:2018ynz}: 
\begin{align}\label{eq:MBparameters}
&m_4=80 \text{ MeV}, \,\, m_5=140 \text{ MeV}, \\\nonumber
&|U_{\mu 5}|^2 = 4/7 \times|U_{\mu 4}|^2 = 1.5 \times 10^{-6} ,\\\nonumber
&|U_{D 4}|^2 = 7/4 \times |U_{D5}|^2 = 7/11 ,\\
&g^\prime = 2,\,\, \chi^2 = 1\times10^{-5}, \text{ and } m_{Z^\prime}=1.25\text{ GeV}.\nonumber
\end{align}
For these parameters, we find that $\nu_4$ is very long lived ($c\tau_4^{0}\simeq 30$ km) and its production only introduces a small deviation from the total number of neutral-current events due to $\nu_\mu\, \mathcal{H} \to \nu_4\, \mathcal{H}$. On the other hand, the signature
\begin{align}
    \nu_\mu\,\, \mathcal{H}\, \to \, & (\nu_5  \,\to\,\nu_4\,\,e^+\,\,e^-)\,\, \mathcal{H}
\end{align}
% %
fakes the MiniBooNE signal due to prompt $\nu_5$ decays ($c\tau_5^{0} \simeq 76$ cm). Decays of the type $\nu_5\to \nu_4 \nu\nu$ are doubly suppressed by the small mixing angles\footnote{In fact, even if the intermediate state becomes lighter, say $m_4 = 50$ MeV, the $\nu_5\to\nu_4\nu_4\nu$ BR would be comparable with that of $\nu_5\to\nu_4 e^+e^-$. Decays of the type $\nu_5\to\nu_4\nu_4 \nu_{(4)}$ dominate whenever $3m_4< m_5$.}, and have the second largest BR of $\simeq 8\times 10^{-5}$. The angular spectrum is analogous to weakly scattering neutrinos due to the large value of $m_{Z^\prime}=1.25$ GeV, but contains only a vectorial piece coupling to matter. We also expect a larger efficiency to select signal events in our model, as we predict larger number of events with overlapping showers (defined approximately as $\Delta \theta_{ee} \lesssim 13^\circ$ at MiniBooNE~\cite{Aguilar-Arevalo:2018gpe}, where $\Delta \theta_{ee}$ is the opening angle of the two electrons), as well as events that are highly asymmetric in energy, both categories faking single electromagnetic shower events. This is due to the low invariant masses $m_{ee}$, that are now bounded by $m_{ee} < m_5-m_4 = 60$ MeV, compared with $m_{ee}<m_4=140$ MeV in Ref.~\cite{Ballett:2018ynz}. For the parameters quoted above, we predict a total number of upscattering events on Carbon plus protons in the full MiniBooNE detector of $\simeq 6.3\times10^3$ for $\nu_4$, and $\simeq 6.1\times10^3$ for $\nu_5$, before detection and signal efficiencies. These numbers correspond to $12.84\times10^{20}$ POT in neutrino mode for $818$~t of CH, and overall efficiencies are expected to be of the order of $5\%$. We note that our rate differs from that of Ref.~\cite{Ballett:2018ynz}, and that the source of discrepancy is still unknown but likely to be due to larger dark couplings than the ones quoted by the authors. The upscattering rate via $Z^\prime$ exchange is proportional to $|\sum_i^3 U_{\mu i}^* U_{Di}|^2 = |U_{\mu 4}^* U_{D4}+U_{\mu 5}^* U_{D5}|^2$ and for our parameters, we find negligible interference with SM bosons.

Beam dump constraints on such unstable $\nu_5$ particle disappear completely and peak searches are weakened, as discussed above. For $\nu_4$, the only relevant constraint is that posed by the beam dump experiment PS-191~\cite{Bernardi:1985ny}, where production through $K^+\to\mu^+\nu_4$ and decay via the $Z^\prime$ enhanced channel $\nu_4\to\nu e^+e^-$ would take place. Although the $|U_{\mu4}|^2$ parameter plays no role in the MiniBooNE explanation, it is heavily correlated with $|U_{\mu5}|^2$ in our three-neutrino theory. Therefore, we can conclude that our benchmark point is most likely in contradiction with the constraint posed by PS-191 on $|U_{\mu4}|^2$, unless such correlation is broken. A more quantitative estimate of the tension is challenging as no event selection information is provided in the original PS-191 analysis and since it assumed only CC decays of heavy neutrinos. We also note that an excess of electron-like events with additional tracks is later reported and attributed to neutrino interactions inside the detector~\cite{Vannucci:1985vs,Bernardi:1986hs,Astier:1989vc} (see also the discussion in Ref.~\cite{Fischer:2019fbw}).

Finally, let us emphasise that the parameters chosen in \refeq{eq:MBparameters} are within the range of mass and mixings required to explain light neutrino masses. For instance, assuming a $2$~GeV scalar, we find that $m_\nu \simeq 0.4$~eV for \refeq{eq:MBparameters}, according to Eq.~(8) of Ref.~\cite{Ballett:2019cqp}. 

\subsection{Impact on dark photon searches} Bounds on the vector portal come from several different sources~\cite{Curtin:2014cca,Bauer:2018onh}. Electroweak precision data and measurements of the $(g-2)$ of the muon and electron constrain our model~\cite{Hook:2010tw}. Major efforts at collider and beam dump experiments led to strong constraints on dark photons by searching for the production and decay of these particles. Such bounds, however, depend on the lifetime of the $Z^\prime$ and on its branching ratio (BR) into charged particles. In our model, the $Z^\prime$ decays invisibly into heavy fermions if $2 m_4<m_{Z^\prime}$ and into light neutrinos otherwise. In the latter case, constraints would be much weaker than usually quoted with only mono-photon searches~\cite{Lees:2017lec} applying. In the former case, however, new signatures arise, where the subsequent decay of $\nu_h$ leads to multi-lepton/multi-meson events, potentially with displaced vertices and providing a very clean experimental signature. Notably, if the $Z^\prime$ decays into $\nu_h$ states that subsequently decay sufficiently fast within the detector, even the ``invisible decay" bounds will be weakened.

\subsubsection{Revisiting $\Delta a_\mu$} The above possibility opens the option to explain the discrepancy between the theoretical prediction~\cite{Davier:2010nc,Davier:2017zfy,Blum:2018mom,Keshavarzi:2018mgv,Davier:2019can} and the experimental value~\cite{Bennett:2006fi} of the $(g-2)$ of the muon via kinetic mixing. For instance, a $1$ GeV $Z^\prime$ with $\chi = 2.2\times10^{-2}$ can explain $a_\mu$. Taking $\nu_4$ around 800 MeV and $m_{Z^\prime}<m_5$, then the $Z^\prime$ would decay into $\nu_4 \nu$ immediately. For the quoted value of the kinetic mixing and $|U_{\mu4}|^2= 10^{-5}$, for instance, the heavy fermions would predominantly decay with sub-cm decay lengths to $e^+ e^-$ and $\mu^+ \mu^-$ pairs plus missing energy, as well as into $\nu \rho^0$ around $1\%$ of the time. This region of the $\chi$ parameter space is constrained only by the BaBar $e^+e^-$ collider searches for visible~\cite{Lees:2014xha} and invisible decays~\cite{Lees:2017lec} of a standard dark photon. Both of these searches would veto the three-body decays of $\nu_4$, opening up a large region of parameter space (see Refs.~\cite{Mohlabeng:2019vrz,Duerr:2019dmv} for a similar discussion in inelastic DM models). Resonance searches still constrain the $Z^\prime$ BR into $e^+ e^-$ and $\mu^+ \mu^-$ which are proportional to $\chi^2$, providing a weak upper bound. A detailed analysis to identify the viable parameter space will be done elsewhere. Note that neutrino masses are too large for the mixing quoted, but can easily be accommodated by extending the neutral fermion sector. 

\subsection{Impact on dark scalar searches}
For the scalar portal, the coupling $\lambda_{\Phi H}$ is rather weakly bound by electroweak precision data and the measurement of the Higgs invisible decay at the level of $\lambda_{\Phi H} \lesssim 0.1$~\cite{Sirunyan:2018owy}. For processes involving $\lambda_{\Phi H}$, the physical observables are suppressed by mass insertions due to the nature of the Higgs interaction. Nevertheless, $\varphi^\prime$ may decay to a pair of $\nu_h$ states and lead to multi-lepton signatures inherited from $\nu_h$ decays, potentially also in the form of displaced vertices. For this reason, bounds on visible decays of $\varphi^\prime$ no longer apply, and the parameter space $(m_{\varphi^\prime},\sin{\theta})$ is wide open when $\varphi^\prime \to \nu_h \nu_h \to$ visible happens in the cm scale.

\subsubsection{Unexplained KOTO events} The new scalar decays can be invoked to explain the recent anomalous number of $K_L\to \pi^0 \nu\overline{\nu}$ events reported by the KOTO collaboration~\cite{KOTO:2019}. In our model, this can be achieved in the range $m_{\varphi^\prime} = 120-170$ MeV and $\theta^2 \simeq 4 \times 10^{-7}$, evading $K^+\to \pi^+\varphi^\prime$ searches at NA62~\cite{NA62:2019} and E949~\cite{Artamonov:2009sz} bounds due to large pion backgrounds at $(p_K-p_\pi)^2\simeq m_\pi^2$~\cite{Hou:2016den}. In addition, if $\varphi^\prime \to \nu_h \nu$ decays are fast, then two scenarios arise. First, if $\nu_4$ is very long lived, \eg, $m_5 \gg m_4 = 100$ MeV, $\chi\to0$, and $|U_{\alpha}|^2 \lesssim 10^{-6}$, then beam dump constraints on $\varphi^\prime$, such as the ones posed by the CHARM experiment~\cite{Bergsma:1985qz}, are relaxed. Secondly, if $\nu_h$ is short-lived, then one may invoke the ``lifetime gap'' explanation proposed in Ref.~\cite{Kitahara:2019lws}. In this case, the dark scalar can decay visibly through a dark cascade with 
\begin{equation}
    c\tau^0(\varphi^\prime\to\nu_4\nu\to\nu\nu Z^\prime\to \nu\nu e^+e^-) \simeq 10 \,\,\text{cm},
\end{equation} 
and so it is vetoed in the large acceptance NA62 and E949 detectors~\cite{NA62:2019}, but would inevitably count as signal for the smaller KOTO detector. The latter scenario is tightly constrained by beam dump searches at $\nu$Cal~\cite{Blumlein:1990ay,Blumlein:2011mv,Blumlein:2013cua}, which, despite their large uncertainties in the region of interest~\cite{Egana-Ugrinovic:2019wzj}, will be mildly alleviated due to the smaller boosts and geometrical acceptance introduced by the dark cascade. This three-portal signature with finite $\nu_4$ lifetime can be achieved for $g^\prime=0.1$, $\chi \simeq 2\times10^{-4}$, $m_{Z^\prime} = 50$ MeV, $m_4=100$ MeV, $|U_{\alpha 4}|^2 \lesssim 10^{-10}$ and $m_{\varphi^\prime}=140$ MeV. 

\subsection{Neutrinophilic limit}
In the limiting case of a neutrinophilic model ($\chi=\lambda_{\Phi H}=0$), the vector and scalar particles present a challenge for detection. Nonetheless, if light, they can be searched for in meson decays~\cite{Laha:2013xua,Bakhti:2017jhm} and at neutrino experiments~\cite{Bakhti:2018avv}.
Finally, the faster decays of $\nu_h$ and its self-interactions can help ameliorate tensions with cosmological observations. We do not comment further on this, but note that great effort has been put into accommodating eV scale sterile neutrinos charged under new forces with cosmological observables~\cite{Hannestad:2013ana,Dasgupta:2013zpn,Mirizzi:2014ama,Chu:2015ipa,Cherry:2016jol,Chu:2018gxk,Song:2018zyl} (see also Ref.~\cite{Escudero:2019gzq} for an interesting discussion where the $Z^\prime$ decay to neutrinos leads to an altered expansions history of the Universe). We note that an eV sterile neutrino with relatively large mixing could be easily accommodated in our ESS framework. The eV neutrino would be mainly in the $\nu_D$ direction and would have strong hidden gauge interactions.

\section{Dark Matter} 
Given the presence of a dark sector, we can ask if the model can accommodate a DM candidate. This can be achieved introducing new fermions that do not mix with the neutrinos, in order to preserve their stability. A minimal solution would be to introduce a fermionic field $\psi_L$ which has $U(1)^\prime$ charge $1/2$. The different charges of $\psi$, $\nu_D$ and $N$ would forbid neutrino mixing. A Majorana mass term 
\begin{equation}
    \psi_L^T C^\dagger \psi_L,
\end{equation} 
would emerge after hidden-symmetry breaking leading to a Majorana DM candidate. Anomaly cancellation requires additional particle content, for instance, promoting all charged fermions to vector-like states.

Another realisation with additional symmetries can be made anomaly free. Following Ref.~\cite{Blennow:2019fhy}, we introduce a pair of chiral fermion fields $\psi_L$ and $\psi_R$, and charge only the latter under the $U(1)^\prime$ symmetry with the same charge as $\nu_D$. This choice ensures anomaly cancellation, and allows us to write
\begin{align}
y_{\psi} \overline{\psi_L} \psi_R \Phi^\dagger,
\end{align}
which after hidden-symmetry breaking yields a Dirac mass $m_\psi$. In general, $\psi_L$ may also have a Majorana mass term $\mu_\psi$, giving rise to a two component Majorana dark matter sector. Setting $\mu_\psi\to0$ is technically natural, and would ensure a Dirac dark matter candidate. In order to avoid $\psi_R-\nu_D$ and $\psi_L-N$ mixing, an additional $\mathbb{Z}_2$ symmetry may be imposed, under which all particles have charge $+1$, except for $\psi_L$ and $\psi_R$, which have charge $-1$. This so-called dark parity can be thought of as a consequence of lepton number $L$. Since $L(\psi)=0$, the Majorana mass term for $N$ is the only soft breaking term, which breaks lepton number by $\Delta L=2$ units. Lepton number is then reduced to conservation of lepton parity $(-1)^L$, which can then be shown to be equivalent to our dark parity~\cite{Ma:2015xla}.

If the scalar and vector portal couplings are small in such scenarios, DM interacts mainly with neutrinos. Direct detection bounds are then evaded, since interactions with matter are loop-suppressed. Indirect detection, on the other hand, is more promising as DM annihilation into neutrinos would dominate. For instance, take the mass of $\psi$ to be smaller than the masses of the $Z^\prime$, $\varphi^\prime$ and of both heavy neutrinos. In this case, the DM annihilation is directly into light neutrinos via $\psi \overline{\psi} \to \nu_i \nu_i$. This yields a mono-energetic neutrino line that can be looked for in large volume neutrino~\cite{Beacom:2006tt,PalomaresRuiz:2007eu} or direct detection experiments~\cite{McKeen:2018pbb}. Alternatively, if $m_\psi$ is larger than the mass of any of our new particles, then the annihilation may be predominantly into such states via $\psi \overline{\psi} \to X X$, where $X=\varphi^\prime, Z^\prime$ or $\nu_h$, which subsequently decay to light neutrinos. In this secluded realisation~\cite{Pospelov:2007mp}, the neutrino spectrum from such annihilation is continuous~\cite{Escudero:2016ksa}, but neutrino-DM interactions are expected to be large and can be searched for in a variety of ways~\cite{Mangano:2006mp,Wilkinson:2014ksa,Farzan:2014gza,Campo:2017nwh,Arguelles:2017atb}.

\section{Conclusions} We have proposed a new model which invokes the existence of a hidden $U(1)^\prime$ symmetry confined to a new dark neutrino sector. Without restricting to the region where our model generates the correct neutrino masses, we explored a series of experimental signatures that may arise from the interplay of portal couplings. The simultaneous presence of neutrino, vector kinetic and scalar mixing in a self-consistent framework allows for a diverse phenomenology. In particular, signatures such as multi-lepton final states with missing energy, displaced vertices, rare leptonic decays, and unique final states in neutrino scattering processes are a hallmark of such non-minimal models. We have also argued that previously excluded parameter space for the minimal realisation of each portal coupling opens up due to faster and semi-visible decays in the dark sector. This also impacts peak searches in meson decays, often regarded as model-independent bounds. For heavy neutrinos that decay to light dark photons, we identify a unique signature at kaon experiments, wherein three charged-lepton final states can be used to search for this class of dark sectors, and in the case of a signal, reveal the heavy neutrino and dark photon masses. This search can be performed at NA62, where significant improvement over current bounds can be achieved. These searches are particularly relevant at low dilepton invariant masses $m_{ee}$, as they could further test the phenomenological model discussed in Refs.~\cite{Bertuzzo:2018itn,Arguelles:2018mtc}, where low-energy di-electrons produced in neutrino upscattering explain the energy distribution of the MiniBooNE excess.

The model also offers exciting new avenues to explain currently-outstanding experimental anomalies. In this article, we discussed how recently-proposed explanations would fit within a non-minimal dark sector like ours, each with their separate choice of parameters. For the MiniBooNE low energy excess, we propose a GeV scale $Z^\prime$ for the angular distribution, and a pair of heavy neutrinos with $m_4\simeq 80$ MeV, and $m_5\simeq140$ MeV. The decays of $\nu_5$ are cm scale, but $\nu_4$ travels much longer distances. Unfortunately, such $\nu_4$ particles have enhanced lifetimes with respect to the minimal model, and are in tension with the PS-191 beam dump experiment unless the correlation present in the three-neutrino model between $\nu_4$ and $\nu_5$ mixing angles is broken. For heavy neutrinos with a larger mass $m_4\simeq 800$ MeV, a $Z^\prime$ of 1 GeV with $\chi=2.2\times10^{-2}$ can explain the discrepancy observed in the anomalous magnetic moment of the muon, $\Delta a_\mu$, while at the same time evading constraints from $Z^\prime \to {\rm invisible}$ at BaBar due to the semi-visible decays of $\nu_4$. Finally, a light scalar with $m_{\varphi^\prime}\simeq 140$ MeV can explain the KOTO anomaly~\cite{Egana-Ugrinovic:2019wzj}. Beam dump bounds on such scalars are completely evaded if it decays to long-lived $\nu_4$ states, or are mildly alleviated if it undergoes a dark cascade to $e^+e^-$ within 10 cm, the latter also evading NA62 bounds from $K^+\to\pi^+\slashed{E}$ searches. A correlation between the $|U_{\alpha 4}|^2$ and $|U_{\alpha 5}|^2$ (as well as between $|U_{D 4}|^2$ and $|U_{D 5}|^2$) neutrino mixing parameters within our three-neutrino model precludes a simultaneous explanation of all the aforementioned anomalies without introducing tensions with current experimental bounds. Nevertheless, the signatures we discussed pertain to a far more general class of models, and deserve further consideration due to their unique character.

\section*{Acknowledgements}
The authors would like to thank Martin Bauer for interesting discussions regarding the dark photon and $(g-2)_\mu$ phenomenology, and Andr\'es Olivares-del-Campo and Arsenii Titov for insights on the dark matter models. We acknowledge interesting discussions with Evgueni Goudzovski on the physics potential of NA62.  This work was partially supported by Conselho Nacional de Ci\^{e}ncia e Tecnologia (CNPq). This project also received support from the European Union's Horizon 2020 research and innovation programme under the Marie Sk\l odowska-Curie grant agreement No. 690575 (RISE InvisiblesPlus) and No. 674896 (ITN Elusives). SP and PB are supported by the European Research Council under ERC Grant NuMass (FP7-IDEAS-ERC ERC-CG 617143). SP acknowledges partial support from the Wolfson Foundation and the Royal Society. The research at the Perimeter Institute is supported in part by the Government of Canada
through NSERC and by the Province of Ontario through
MEDT.

\appendix

\section{Kinematics of exotic kaon decays}
\begin{figure*}[t]
\centering
\includegraphics[width=0.49\textwidth]{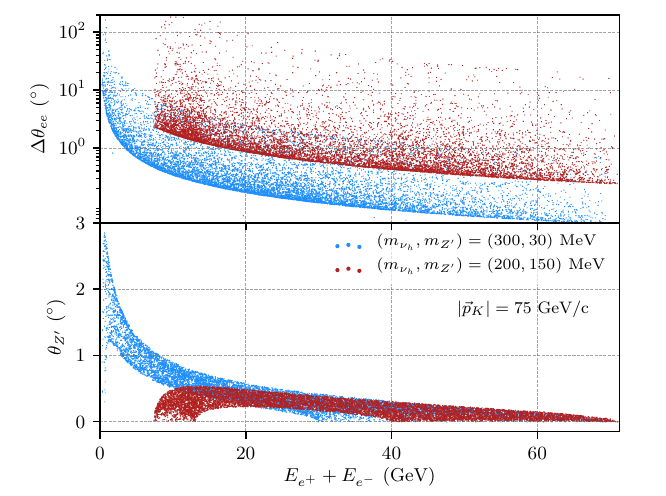}
\includegraphics[width=0.49\textwidth]{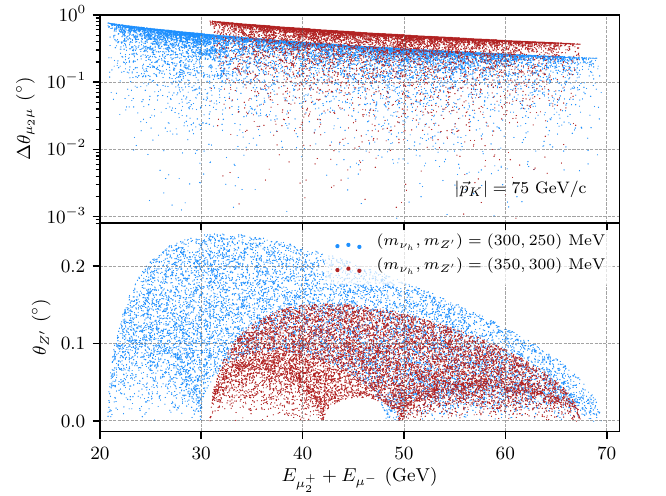}
\includegraphics[width=0.49\textwidth]{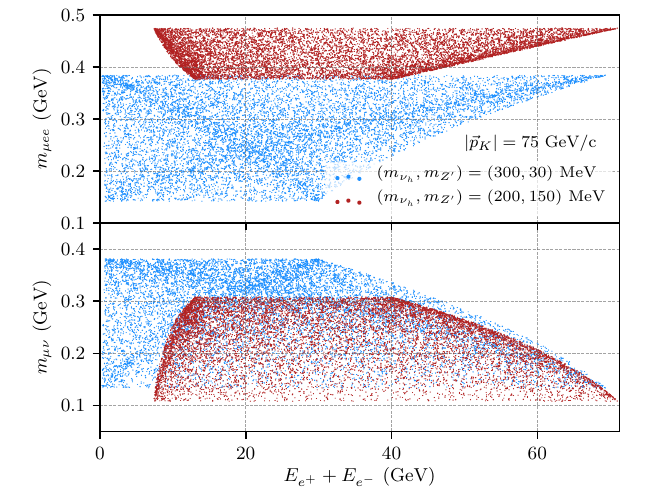}
\includegraphics[width=0.49\textwidth]{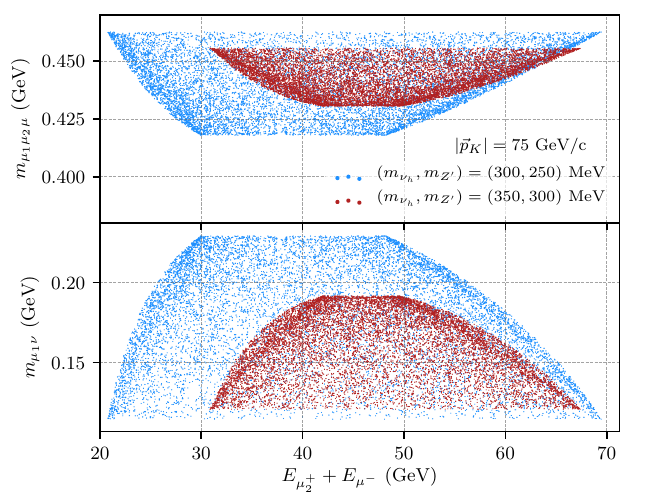}    
\caption{Kinematics of our $K^+\to\mu^+\nu \, e^+e^-$ (left column) and $K^+\to\mu^+_1\nu\, \mu^+_2\mu^-$ (right column) events at kaon energies relevant to NA62. We plot $10^4$ events in true Monte Carlo variables in each panel. In the top row we show both the separation angle between the decay products of the $Z^\prime$, $\Delta \theta_{\beta\beta}$, and the angle of the $Z^\prime$ momentum with respect to the kaon beam, $\theta_{Z^\prime}$, both versus the total energy of the decay products of the $Z^\prime$, $E_{\beta^+}+E_{\beta^-}$. In the bottom row we show the invariant masses $m_{\mu_{(1)}\beta\beta}=\sqrt{(p_{\mu_{(1)}}+p_{\beta} + p_\beta)^2}$ and $m_{\mu_{(1)}\nu}=\sqrt{(p_{\mu_{(1})}+p_\nu)^2}$, also versus the total energy $E_{\beta^+}+E_{\beta^-}$.  \label{fig:signalkinematics}}
\end{figure*}

In this appendix we expand on details of the rare kaon decay searches we propose. Beyond the peaks in $m_{\beta\beta}=m_{Z^\prime}$, $m_{p_K-p_\alpha}=m_{\nu_h}$, and $m_{\rm miss} = 0$, one may explore the correlation and absolute range of two other relevant invariant masses, namely
\begin{align}
    m_{\alpha\nu}^2&=(p_\mu + p_\nu)^2=(p_K-p_{\beta^+}-p_{\beta^-})^2,
    \\
m_{\alpha \beta \beta}^2&=(p_\alpha + p_{\beta^+} + p_{\beta^-})^2.
\end{align}
Their absolute kinematical range is determined, and shown here for the reader's convenience
\begin{widetext}
\begin{align}
    \left(m_{\alpha\nu}^2\right)_{\rm min}^{\rm max} &= m_K^2\Bigg[ 1 + x_{Z^\prime} - \frac{(1-x_{\alpha}+x_{h})(1 + y) \mp (1 - y) \lambda^{1/2}(1,x_h,x_{\alpha})}{2} \Bigg],
    \\
    \left(m_{\alpha \beta \beta}^2\right)_{\rm min}^{\rm max} &= m_K^2\left[ 1 - (1 - y)\frac{(1-x_{\alpha}+x_{h}) \mp \lambda^{1/2}(1,x_h,x_{\alpha})}{2} \right],
\end{align}
\end{widetext}
where $\lambda(a,b,c)=(a-b-c)^2-4bc$ is the K\"allen function, $x_{Z^\prime}=(m_{Z^\prime}/m_{K})^2$, $x_{h}=(m_{\nu_h}/m_{K})^2$, $x_{\alpha}=(m_{\alpha}/m_{K})^2$, and $y=(m_{Z^\prime}/m_{\nu_h})^2$.

To illustrate the properties of our signal, we develop our own toy Monte Carlo simulation, and generate $10^4$ events for $K^+\to \mu^+(\nu_h\to\nu(Z^\prime\to e^+e^-))$ and $K^+\to \mu_1^+(\nu_h\to\nu(Z^\prime\to \mu_2^+\mu^-))$ decays in a few cases of interest. We fix the kaon momentum to the average value in the NA62 beam, namely $75$ GeV/c. We plot the invariant masses $m_{\nu\alpha}$ and $m_{\alpha\beta\beta}$, as well as the opening angle between the leptons $\Delta \theta_{\beta \beta}$, and the angle of the $Z^\prime$ momentum with respect to the Kaon beam, $\theta_{Z^\prime}$, versus total $Z^\prime$ energy in \reffig{fig:signalkinematics}. As expected, the kinematics of the signal is highly boosted, with very small angles between the decay products. Another feature is the broad distributions in invariant masses, except for $m_{\mu_1\mu_2\mu}$, which has somewhat of a smaller range due to the limited phase space of each decay reaction.

%%%%%%%%%%%%%%%%%%%%%%%%%%%%%%%%%%%%%%%%%
\bibliographystyle{apsrev4-1}
\bibliography{main}{}
%%%%%%%%%%%%%%%%%%%%%%%%%%%%%%%%%%%%%%%%%

\end{document}